\documentclass[reprint,aps,sort&compress,amsmath,amssymb,superscriptaddress]{revtex4-1}
\usepackage{bm}
\usepackage{graphicx}
\usepackage{units}
\usepackage[english]{babel}
\usepackage[colorlinks, citecolor=blue, urlcolor=blue, linkcolor=red,breaklinks]{hyperref}
\begin{document}

\title{Entanglement harvesting in buckled honeycomb lattices by vacuum fluctuations in a microcavity}

\author{Facundo Arreyes}
\affiliation{Departamento de F\'isica, Universidad Nacional del Sur, Av. Alem 1253, B8000, Bah\'ia Blanca, Argentina}
\author{Federico Escudero}
\affiliation{Departamento de F\'isica, Universidad Nacional del Sur, Av. Alem 1253, B8000, Bah\'ia Blanca, Argentina}
\affiliation{Instituto de F\'isica del Sur, Conicet, Av. Alem 1253, B8000, Bah\'ia Blanca, Argentina}
\author{Juan Sebastián Ardenghi}
\email{jsardenghi@gmail.com}
\affiliation{Departamento de F\'isica, Universidad Nacional del Sur, Av. Alem 1253, B8000, Bah\'ia Blanca, Argentina}
\affiliation{Instituto de F\'isica del Sur, Conicet, Av. Alem 1253, B8000, Bah\'ia Blanca, Argentina}
\author{Alfredo Juan}
\affiliation{Departamento de F\'isica, Universidad Nacional del Sur, Av. Alem 1253, B8000, Bah\'ia Blanca, Argentina}
\affiliation{Instituto de F\'isica del Sur, Conicet, Av. Alem 1253, B8000, Bah\'ia Blanca, Argentina}

\begin{abstract}
	We study the entanglement harvesting between two identical buckled honeycomb lattices placed inside a planar microcavity. By applying time dependent perturbation theory, we obtain quantum correlations between both layers induced by the cavity field. Considering the vacuum state as the initial state of the cavity field and tracing out the time-evolved degrees of freedom, we analyze the entanglement formation using the concurrence measure. We show that the concurrence depends on the virtual photon exchanged and the positions of the layer through the interlayer photon propagator. Furthermore, we find that the formation of entanglement between equal energy electrons tends to be enhanced when they move in perpendicular directions. Our results indicate that a buckled honeycomb structure and a large spin-orbit interaction favor the entanglement harvesting.

\end{abstract}

\maketitle

%\onecolumngrid

\section{Introduction}

It is well known that quantum particles with finite spectrum can be used as probes to measure the entangled vacuum state of a quantum field \cite{reznik2}. Since the ground state of a quantum field is highly entangled in space, it allows correlations of different observables at space-like distances \cite{valentini,resnik,pozak}. These correlations can be harvested through the time-evolved quantum states of the detectors, which is known as entanglement harvesting \cite{resnik,valentini}. Scalar fields have been widely studied \cite{pozak} in this context and applied for entanglement farming \cite{mar1}, metrology \cite{salt}, cosmology \cite{stee,mar2,mar3}, quantum information \cite{jon,haw,al}, quantum energy teleportation \cite{hotta} and entanglement between real and virtual particles \cite{jsa1,jsaA}. Experimental implementations of the Unruh-DeWitt model (UDW), in which the most simplest is a linear coupling of a pointlike two-level system with a quantum field, have been developed in atomic systems where atoms are considered the first quantized system that interacts with the second quantized electromagnetic field \cite{ols1,ols2}. Generalizations to spatial smeared detectors have been studied in \cite{strit}, where a spatial smearing function is included in order to allow the two level system to have a finite extension in space. The spread of the quantum center of mass wave function affects the ability of the detectors to get entangled with each other.

When space is confined, for example, by placing quantum detectors inside a microcavity, the coupling between photons and matter can be enhanced \cite{li}, allowing the formation of polariton states \cite{byrnes,Ysun}. Due to the cavity electromagnetic field, vacuum fluctuations give rise to virtual photons that do not require an active external field pumping and can induce appreciable modifications in the electronic properties of the combined system \cite{marino,ribeiro,escudero0}. In particular, previous research \cite{stock,frisk} has demonstrated a substantial reduction in the effective electron-photon interaction volume in nanoplasmonic cavities, reaching orders of magnitude below the free space diffraction limit \cite{Xliu, maissen}. These enhanced vacuum fluctuations may induce various effects, including superconductive states in 2D systems \cite{schla,cotlet,laus}, carrier density-dependent Mott insulator states \cite{Ycao,Ycao2}, and Fermi velocity renormalization in monolayers \cite{balleste,gupta,escudero}.
Interestingly, two-dimensional systems are suitable for mimicking the Unruh-DeWitt interaction through the minimal coupling between the sublattice base and the cavity field. In \cite{jsa2018} and \cite{facu}, by neglecting tunneling and electron-electron interactions in different layers, entanglement harvesting and non-classical correlations in double-layer graphene in a microcavity were obtained by means of exchanging virtual photons. The study of light-matter interaction in two-dimensional structures has become relevant for quantum optical devices and has branched into different applications, such as band gap opening \cite{kib3,kib,kib2,kristi}, fractional quantum Hall effect \cite{jia}, excitonic effects \cite{khari,sodeman}, Casimir forces \cite{bost,been,dros,zare} and effective interactions that renormalize the free Hamiltonian parameters \cite{jsa2}. 

In recent years, new 2D materials have been experimentally obtained, such as molybdenum disulfide (MoS$_{2}$) \cite{chu}, silicene \cite{spen}, germanene \cite{gill,bian}, phosphorene \cite{yar11}, transition metal dichalcogenides and hexagonal boron nitride \cite{roldan}, and two-dimensional SiC monolayers \cite{yar18}, which are similar to graphene but with buckled geometry and different atoms at each lattice site. Among these materials, the buckled honeycomb Dirac systems are particularly interesting due to the vertical displacement between the two sublattices, as well as a spin-orbit interaction (SOI) greater than in graphene \cite{liu,yao}. This allows one to open a tunable bandgap by applying an external electric field \cite{ezawa}, making those systems suitable for their use as non-localized quantum detectors exchanging virtual photons. Thus, the aim of this work is to study the entanglement harvesting between two buckled honeycomb lattices placed inside a microcavity, and analyze how the SOI affects the entanglement formation with respect to graphene systems.
By means of time-dependent perturbation theory and the concurrence measure, we find that a buckled honeycomb structure and a non-negligible SOI actually favor the entanglement formation between both layers inside the microcavity.  

The remainder of this paper is organized as follows. In Sect. \ref{sec:Tmodel} we introduce the formalism to compute the time-dependent perturbation theory. In Sect. \ref{sec:RandD} we present our results and discussions for different electronic states in both buckled monolayers. The concurrence and entanglement entropy are computed for different initial energies, angles, and SOI. Our conclusion follows in Sec. \ref{sec:Conclusions}.

\section{Theoretical model}\label{sec:Tmodel}

\subsection{Cavity-mediated interactions}

We model the electronic properties of the buckled honeycomb layers within the low-energy Dirac approximation, i.e., where electrons behave as relativistic fermions with an effective Fermi velocity of about $v_F\sim c/300$ \cite{spen,ezawa}. The Hamiltonian of two buckled honeycomb layers coupled to the electromagnetic field of the cavity is given by $H=H_{0}+H_{F}+H_{int}$, where
\begin{align}
	H_{0} & =\sum\limits _{i=1,2}(v_{F}\boldsymbol{\sigma}_{i}\cdot\mathbf{p}_{i}+\sigma_{i,z}\Delta_{s_{i}\eta_{i}}),\\
	H_{F} & =\underset{n,\mathbf{q},\lambda}{\sum}\hbar\omega_{n,\mathbf{q},\lambda}a_{n,\mathbf{q},\lambda}^{\dagger}a_{n,\mathbf{q},\lambda},\\
	H_{int} & =-ev_{F}\sum\limits _{i=1,2}\boldsymbol{\sigma}_{i}\cdot\mathbf{A}_{i}.
\end{align}
Here, $H_{0}$ is the low-energy Hamiltonian of the two uncoupled buckled honeycomb lattices \cite{peres-guinea,spen}. The index $i$ runs over the two electrons in different layers, with $\boldsymbol{\sigma}_{i}=\left(\sigma_{i,x},\sigma_{i,y}\right)$ and $\sigma_{i,z}$ being the Pauli matrices acting on its corresponding sublattice space \cite{kristi,spen,peres-guinea}. The Fermi velocity $v_{F}$ depends on the two-dimensional material (e.g., $v_{F}=5.5\times10^{5}$ m/s in silicene \cite{liu}). The coefficient of the diagonal term in $H_{0}$ is given by $\Delta_{s_{i}\eta_{i}}=\eta_{i}s_{i}\lambda_{so}-\ell E_{z}$, where $s_i,\eta_i=\pm$ are the spin and valley indices, $\lambda_{so}$ is the spin-orbit coupling strength, $\ell$ is the separation between the $A$ and $B$ sublattices, and $E_{z}$ is a perpendicular electric field that can be applied to modulate the band gap. In what follows, we consider, for simplicity, no external perpendicular electric field. Since the two valleys are separated by a large momentum gap, interval scattering is generally negligible at low energies \cite{par,zare1}. Therefore, in our analysis, we will focus on a single valley and for convenience we will take $\eta=1$. 

\begin{figure}[t]
	\includegraphics[scale=0.63]{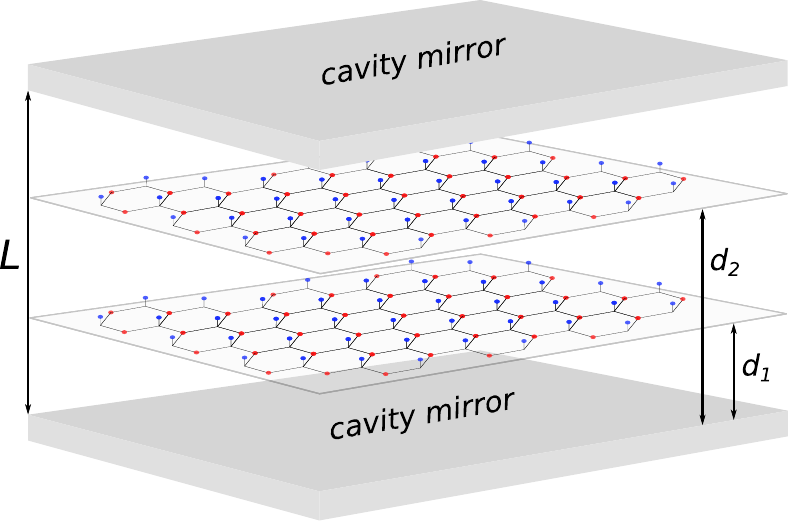}
	\caption{Schematic setup of the microcavity of length $L$, with two identical buckled honeycomb lattices placed at distances $d_{1}$ and $d_{2}$.}
	\label{micro}
\end{figure}

The Hamiltonian $H_{F}$ describes the electromagnetic field in the cavity. The corresponding photon frequencies read \cite{kakazu}
\begin{equation}
\omega_{n,\mathbf{q}}=c\sqrt{\mathbf{q}^{2}+\left(\frac{n\pi}{L}\right)^{2}},
\end{equation}
where $c$ is the speed of light inside the cavity, $\mathbf{q}$ in the in-plane photon momentum and $L$ is the cavity length (see Fig. \ref{micro}). Note that in our configuration, the photon momentum is quantized along the cavity length as $q_z=n\pi/L$, where $n$ is an integer accounting for the standing wave modes inside the cavity \cite{kakazu}. The operators $a_{n,\mathbf{q},\lambda}^{\dagger}$ and $\:a_{n,\mathbf{q},\lambda}$ are the creation and annihilation operators of a cavity photon in the $n$ mode, with momentum $\mathbf{q }$ and polarization $\lambda$; they obey the usual commutation relations $\left[a_{n,\mathbf{q},\lambda},a_{n',\mathbf{q}',\lambda'}^{\dagger}\right]=\delta_{n,n^{\prime}}\delta_{\lambda,\lambda^{\prime}}\delta_{\mathbf{qq}^{\prime}}$ \cite{kib}.

The Hamiltonian $H_{int}$ describes, within the minimal coupling, the interaction between the electromagnetic field in the cavity and the electrons in the buckled honeycomb lattices. The quantization of the potential vector inside the cavity can be written in terms of the creation and annihilation operators for each mode as \cite{jsa2018,jsa2}%
\begin{equation}
	\mathbf{A}_{i}\mathbf{=}\underset{n,\mathbf{q,\lambda}}{\sum}\frac{\gamma}{\sqrt{S\omega_{n,\mathbf{q}}}}\sin\left(\frac{n\pi d_{i}}{L}\right)\mathbf{e}_{\lambda}a_{n,\mathbf{q},\lambda}e^{i\left(\mathbf{q\cdot\mathbf{r}}_{i}\mathbf{-\omega}_{n,\mathbf{q}}t\right)}+h.c.,
\end{equation}
where $\mathbf{e}_{\lambda}=\left(\mathbf{e}_{x}+i\lambda\mathbf{e}_{y}\right)/\sqrt{2}$ are the circular polarization directions (with $\lambda= \pm $), $d_i$ is the position of the $i$ layer with respect to the bottom cavity mirror (cf. Fig. \ref{micro}), $S$ is the area of the system, and $\gamma=\sqrt{\hbar/\varepsilon_0 \varepsilon_r L}$. 

Electronic hopping between layers can be neglected provided that the interlayer separation $\left\vert d_{1}-d_{2}\right\vert $ is larger than a characteristic distance $l$ between charge carriers in each layer \cite{pla,gor,fedeprb}. For simplicity we write $\Delta
_{s_{i}}=\Delta _{s_{i}\eta _{i}=+1}=s_{i}\lambda _{so}$ for one valley in $H_{0}$. The eigenstates in each layer can be obtained in terms of the sublattice basis as \cite{jsamag}
\begin{equation}
\left\vert  \nu,s\right\rangle =\frac{1}{\sqrt{1+\chi_{k,\nu,s}^{2}}}\left[\left\vert A,s\right\rangle +\chi_{k,\nu,s}e^{i\phi_{\mathbf{k}}}\left\vert B,s\right\rangle \right].\label{3a}
\end{equation}
where $\nu=\pm$ is the band index and $\phi_{\mathbf{k}}=\arctan\left(k_{y}/k_{x}\right)$.
The function $\chi_{k,\nu,s}$ reads
\begin{equation}
\chi_{k,\nu,s}=\epsilon_{\mathbf{k}}^{-1}\left(E_{\mathbf{k},\nu,s}-\Delta_{s}\right),\label{eq:chi}
\end{equation}
where $E_{\mathbf{k},\nu,s}=\nu\sqrt{\Delta_{s}^{2}+\epsilon_{\mathbf{k}}^{2}}$,
with $\epsilon_{\mathbf{k}}=\hbar v_{F}\left|\mathbf{k}\right|$,
are the eigenvalues of $H_{0}$ in each layer.

It is convenient to write the interaction between the monolayers and the electromagnetic field as 
\begin{align}
	H_{int} & =-ev_{F}\sum\limits _{i=1,2}\sum\limits _{\lambda=\pm}\sigma_{\lambda}^{i}A_{\lambda}^{i}\left(\mathbf{r}_{i},t\right), \label{4a1}\\
	A_{\lambda}^{i}\left(\mathbf{r}_{i},t\right) & =\underset{n,\mathbf{q}}{\sum}\frac{\gamma}{\sqrt{S\omega_{n,\mathbf{q}}}}\sin\left(\frac{n\pi d_{i}}{L}\right)\left[a_{n,\mathbf{q},\lambda}e^{i\left(\mathbf{q\cdot\mathbf{r}}_{i}\mathbf{-\omega}_{n,\mathbf{q}}t\right)}\right.\nonumber \\
	& \left.\qquad\qquad+\,a_{n,\mathbf{q},-\lambda}^{\dagger}e^{-i\left(\mathbf{q\cdot\mathbf{r}}_{i}\mathbf{-\omega}_{n,\mathbf{q}}t\right)}\right]. \label{A}
\end{align}
where we have used $\sigma_{\lambda}=\left(\sigma_{x}+i\lambda\sigma_{y}\right)/2$. Eq. (\ref{4a1}) is more suitable for the application of the perturbation expansion. The coupling between the sublattice operators $\sigma_{\lambda}$ with $a$ and $a^{\dagger }$ involve the processes where an electron tunnels from one sublattice to the other through the absorption or emission of a photon. An interesting process is one in which the photon emitted by one layer is absorbed by the other, as it implies an interaction between electrons in different layers through the exchange of a virtual photon \cite{mar3,marti,leon}.

\subsection{Time-dependent perturbation theory}

In this work our interest is to study the cavity-mediated entanglement between two buckled honeycomb lattices. Within the regime  $\epsilon_{\mathbf{k}}\gg\lambda_{so}$, the ratio between the kinetic ($\left[H_{0}\right]\sim E_{\mathbf{k}}$) and interaction ($\left[H_{int}\right]\sim\gamma\sqrt{S\omega_{0}}$) energies scales as
\begin{equation}
R=\frac{\left[H_{int}\right]}{\left[H_{0}\right]}\sim\sqrt{\frac{4\alpha}{S\varepsilon_{r}}}\frac{1}{k},
\end{equation}
where $k$ is a characteristic momentum for the kinetic energy $\epsilon_{\mathbf{k}}\sim\hbar v_{F}k$, and $\alpha=e^{2}/4\pi\varepsilon_{0}\hbar c\sim1/137$ is the fine structure constant. For typical values $\epsilon_{\mathbf{k}}\sim0.1\,\mathrm{eV}$ (corresponding to electronic carrier densities $n_{e}\sim10^{12}\,\mathrm{cm^{-2}}$), monolayer dimensions, and dielectric constants $\varepsilon_{r}\sim4$, one has $R\sim10^{-4}$. Thus cavity-mediated interaction are weak compared to the kinetic contribution \cite{kib,kib2}. In what follows we then consider the interaction $H_{int}$ as a perturbation to the Hamiltonian $H_{0}+H_{F}$.

Methods for finding the time-evolved state are straightforward and can be found throughout the literature \cite{pozak}. For simplicity, we present the most general and important results. Applying time-dependent perturbation theory, the time evolution operator $U$ for the full system can be expanded as $U=U_{0}+U_{1}+U_{2}+...$, where in general
\begin{align}
	U_{n}\left(t\right) & =\frac{1}{n!}\int_{0}^{t}dt_{1}\ldots\int_{0}^{t}dt_{n}\nonumber \\
	& \quad\times T_{t}\left[H_{int}\left(t_{1}\right)\ldots H_{int}\left(t_{n}\right)\right].
\end{align}
Here $T_t$ is the time-ordering operator, and $H_{int}\left(t\right)$ is the interacting Hamiltonian in the interaction picture: 
\begin{equation}
	H_{int}\left(t\right)=e^{i\left(H_{0}+H_{F}\right)t/\hbar}H_{int}e^{-i\left(H_{0}+H_{F}\right)t/\hbar}
\end{equation}
Given an initial density matrix $\rho _{0}$, the final density matrix $\rho _{T}$ is then given by
\begin{equation}
	\rho_{T}=U\rho_{0}U^{\dagger}=\left[I+U_{1}+U_{2}+\cdots\right]\rho_{0}[I+U_{1}^{\dagger}+U_{2}^{\dagger}+\cdots].  \label{a9}
\end{equation}
Assuming small perturbations, we take $\rho_{T}\simeq\rho_{0}+\rho_{1}+\rho_{2}$, where
\begin{align}
	\rho_{1} & =U_{1}\rho_{0}+\rho_{0}U_{1}^{\dagger},\\
	\rho_{2} & =U_{2}\rho_{0}+U_{1}\rho_{0}U_{1}^{\dagger}+\rho_{0}U_{2}^{\dagger}.
\end{align}

We shall consider that the initial state of the monolayers and the
electromagnetic field is
\begin{equation}
	\rho _{0}=\left\vert \Omega _{0}\right\rangle \left\langle \Omega
	_{0}\right\vert \otimes \rho _{S},  \label{a11}
\end{equation}
\smallskip 
where $\left\vert \Omega_{0}\right\rangle =\left\vert \Omega_{0}^{+},\Omega_{0}^{-}\right\rangle$ is the vacuum state of the electromagnetic field with circular polarization $\pm $, and 
\begin{equation}
\rho_{S}=\left|  \nu_{1},s_{1}\right\rangle \left\langle \nu_{1},s_{1}\right|\otimes\left| \nu_{2},s_{2}\right\rangle \left\langle \nu_{2},s_{2}\right| \label{ro}
\end{equation}
is the initial density matrix of the electronic system.

We are now interested in the partial state of the
electrons in the monolayers after the interaction with the quantum field, which is given by the trace
\begin{equation}
	\rho (t)=T_{\mathrm{r}}(U\rho _{0}U^{\dagger}).  \label{a12}
\end{equation}%
This means that the non-diagonal terms in the field, arising by time evolution, will not contribute to $\rho (t)$. To trace out the field degrees of freedom implies that we are not measuring the field and, consequently, that any field excitation is virtual. Then it is not difficult to show that up to second order $\rho (t)$ evolves as
\begin{align}
	\rho\left(t\right) & =\rho_{S}-\left(\frac{ev_{F}}{\hbar}\right)^{2}\sum\limits_{i,\lambda}\sum\limits_{j,\lambda^{\prime}}\int_{0}^{t}dt_{1}\nonumber\nonumber \\
	& \times\left[\int_{0}^{t_{1}}dt_{2}\mathcal{D}_{ij}^{\lambda,\lambda'}\left(\mathbf{r}_{i},\mathbf{r}_{j},t_{1},t_{2}\right)\rho_{S}\sigma_{\lambda}^{i}\left(t_{1}\right) \sigma_{\lambda^{\prime}}^{j}\left(t_{2}\right)\right.\nonumber\nonumber \\
	& +\int_{0}^{t_{1}}dt_{2}\mathcal{D}_{ij}^{\lambda,\lambda'}\left(\mathbf{r}_{i},\mathbf{r}_{j},t_{1},t_{2}\right)\sigma_{\lambda}^{i}\left(t_{1}\right)\sigma_{\lambda^{\prime}}^{j}\left(t_{2}\right)\rho_{S}\nonumber\nonumber \\
	& \left.-\int_{0}^{t}dt_{2}\mathcal{D}_{ij}^{\lambda,\lambda'}\left(\mathbf{r}_{i},\mathbf{r}_{j},t_{1},t_{2}\right)\sigma_{\lambda}^{i}\left(t_{1}\right)\rho_{S}\sigma_{\lambda^{\prime}}^{j}\left(t_{2}\right)\right],\label{aa13}
\end{align}
where
\begin{equation}\label{propagator}
\mathcal{D}_{ij}^{\lambda,\lambda'}\left(\mathbf{r}_{i},\mathbf{r}_{j},t_{1},t_{2}\right)=\left\langle \Omega_{0}\right\vert A_{\lambda}^{i}\left(\mathbf{r}_{i},t_{1}\right)A_{\lambda^{\prime}}^{j}\left(\mathbf{r}_{j},t_{2}\right)\left\vert \Omega_{0}\right\rangle 
\end{equation}
is the photon propagator in space-time coordinates, and
\begin{align}
%A_{\lambda}^{i}\left(\mathbf{r},t\right) & =e^{iH_{F}t/\hbar}A_{\lambda}^{i}e^{-iH_{F}t/\hbar},\\
\sigma_{\lambda}^{i}\left(t\right) & =e^{iH_{0}t/\hbar}\sigma_{\lambda}^{i}e^{-iH_{0}t/\hbar}
\end{align}
is the time evolved sublattice operator of electrons. The above equation can be compared with Eq. (9) of \cite{marti} by computing the matrix elements of $\rho (t)$ as $\left\langle 
\mathbf{k}_{1}^{\prime }\mathbf{,k}_{2}^{\prime }\right\vert \rho
(t)\left\vert \mathbf{k}_{1},\mathbf{k}_{2}\right\rangle$. Indeed, by introducing the resolution of identity in the real representation as $I=\int d^{2}\mathbf{r}_{1}d^{2}\mathbf{r}_{2}\left\vert \mathbf{r}_{1},\mathbf{r}_{2}\right\rangle \left\langle\mathbf{r}_{1},\mathbf{r}_{2}\right\vert $, from Eq. \eqref{aa13} we get

\begin{widetext}
\begin{align}
\left\langle \mathbf{k}_{1}^{\prime}\mathbf{,k}_{2}^{\prime}\right\vert \rho\left(t\right) & \left\vert \mathbf{k}_{1},\mathbf{k}_{2}\right\rangle =\int d^{2}\mathbf{r}_{1}d^{2}\mathbf{r}_{2}\left\{ e^{-i(\mathbf{k}_{1}^{\prime}-\mathbf{k}_{1})\cdot\mathbf{r}_{1}}e^{-i(\mathbf{k}_{2}^{\prime}-\mathbf{k}_{2})\cdot\mathbf{r}_{2}}\rho_{S}\right.-\left(\frac{ev_{F}}{\hbar}\right)^{2}\sum\limits_{i,\lambda}\sum\limits_{j,\lambda^{\prime}}\int_{0}^{t}dt_{1}\nonumber\nonumber \\
\times & \left[\int_{0}^{t_{1}}dt_{2}W_{ij}^{\lambda,\lambda^{\prime}}\sigma_{\lambda}^{i}\left(t_{1}\right)\sigma_{\lambda^{\prime}}^{j}\left(t_{2}\right)\rho_{S}+\int_{0}^{t_{1}}dt_{2}W_{ij}^{\lambda,\lambda^{\prime}}\rho_{S}\sigma_{\lambda}^{i}\left(t_{2}\right)\sigma_{\lambda^{\prime}}^{j}\left(t_{1}\right)-\int_{0}^{t}dt_{2}W_{ij}^{\lambda,\lambda^{\prime}}\sigma_{\lambda}^{i}\left(t_{1}\right)\rho_{S}\sigma_{\lambda^{\prime}}^{j}\left(t_{2}\right)\right],\label{eq:rhoK}
\end{align}
\end{widetext}
where 
\begin{equation}
	W_{ij}^{\lambda,\lambda^{\prime}}=\mathcal{D}_{ij}^{\lambda,\lambda'}e^{-i\left(\mathbf{k}_{1}^{\prime}-\mathbf{k}_{1}\right)\cdot\mathbf{r}_{1}}e^{-i\left(\mathbf{k}_{2}^{\prime}-\mathbf{k}_{2}\right)\cdot\mathbf{r}_{2}},  \label{aa13.2}
\end{equation}
which is the analogous of the pullback of the Wightman function on the layers smeared wordlines (see Eq. (10) of \cite{marti}). Basically it is defined by the plane waves of the initial states multiplied by the photon propagator, which, when integrated in space, gives the Fourier transform of the propagator. In this case, the profile function of the smearing is the plane waves $e^{-i(\mathbf{k}_{i}^{\prime }-\mathbf{k}_{i})\cdot \mathbf{r}_{i}}$ in each  layer, obtained from the real representation of the initial and final states. For simplicity we are considering plane wave solutions for both layers, but more general in and out states can be considered, such as wave packets with compact smearing functions. The analogous to the monopole detector of Eq. (4) in \cite{marti} are
the time evolved operators $\sigma_{\lambda}^{i}\left(t\right)$ (see \cite{jsa2})
\begin{equation}
\sigma_{\lambda}\left(t\right)=\frac{e^{i\lambda\phi_{\mathbf{k}}}}{\Delta\chi^{2}} \left( \Gamma_{k,s}^{z} \tau_z +\sum_{\nu=\pm}\Gamma_{\lambda,k,s}^{\nu}\tau_{\nu}\right),\label{aa13.3}
\end{equation}
where $\tau _{z}=\left\vert -,s\right\rangle \left\langle -,s\right\vert - \left\vert +,s\right\rangle \left\langle +,s\right\vert $, $\tau_{\nu}=\left|\nu,s\right\rangle \left\langle -\nu,s\right|$ and
\begin{align}
\Delta \chi  &=\chi _{k,+,s}-\chi _{k,-,s},\\
\Gamma_{k,s}^{z} & =\left(1+\chi_{k,+,s}^{2}\right)\chi_{k,-,s},\\
\Gamma_{\lambda,k,s}^{\nu} & =e^{2iE_{\mathbf{k},\nu,s}t/\hbar}\sqrt{\left(1+\chi_{k,+,s}^{2}\right)\left(1+\chi_{k,-,s}^{2}\right)}\chi_{k,-\lambda\nu,s},
\end{align}
The operator $\sigma _{\lambda }(t)$ is a linear combination of the operators \( \tau_z \) and \( \tau_\pm \). The operator $\tau_z$ consistently generates a time-independent antisymmetric superposition in the conduction and valence band states. Simultaneously, the operators $\tau_{\pm}$ induce interband transitions which fluctuate in time due to the exponential dependence on the electron's energy. Since, in the end, we only consider cases where \(i \neq j\), the action of \(\sigma_\lambda\) amounts to applying the aforementioned procedure separately to state 1 and state 2.

Replacing Eq. (\ref{aa13.3}) in Eq. (\ref{eq:rhoK}) yields the quantum processes analogous to the monopole detector, where a new term $\tau _{z}$ appears and implies that there can be emission and absorption of photons without electronic transition between bands. An important difference from our work and the detector formalism used in \cite{marti} is the spatial profile of the detector, since in our model electrons are delocalized across the entire surface of the layer, unlike the Gaussian smearing function used in \cite{marti}. In turn, the field propagator is computed by performing the integration in the field modes, while the propagator, obtained in coordinate space, is convoluted with the spatial profile of the detectors. 

In this work, we could proceed in a similar manner but is not necessary to perform the field modes integration. Indeed, the plane wave profile of the layer detectors allows us to integrate first in $d^{2}\mathbf{r}_{1}$ and $d^{2}\mathbf{r}_{2}$, and combine the plane wave profiles with $e^{\pm\mathbf{q}\cdot\mathbf{r}}$ to obtain momentum conservation. This leads to
\begin{align}
	\int d^{2}\mathbf{r}_{1}d^{2}\mathbf{r}_{2}W_{ij}^{\lambda,\lambda^{\prime}} & =\delta_{\lambda,-\lambda^{\prime}}{\underset{n,\mathbf{q}}{\sum}}\gamma^{2}\mathbf{\sin}\left(\frac{\pi nd_{i}}{L}\right)\sin\left(\frac{\pi nd_{j}}{L}\right)\nonumber \\
	& \quad\times\frac{e^{-i\omega_{n,\mathbf{q}} \left(t_{1}-t_{2}\right)}}{\omega_{n,\mathbf{q}}} \tilde{\delta}_{ij} .
\end{align}
Here we defined the general delta function
\begin{equation}
	\tilde{\delta}_{ij}=\delta_{ij} \delta_{\mathbf{k}_{1}^{\prime},\mathbf{k}_{1}} \delta_{\mathbf{k}_{2}^{\prime},\mathbf{k}_{2}}+(1-\delta_{ij}) \delta_{\mathbf{k}_{1}^{\prime},\mathbf{k}_{1}+\mathbf{q}} \delta_{\mathbf{k}_{2}^{\prime},\mathbf{k}_{2}-\mathbf{q}},
\end{equation}
which accounts for the case $i=j$ where it can only be $\mathbf{k'}_1=\mathbf{k}_1$ and $\mathbf{k'}_2=\mathbf{k}_2$, and the case $i\neq j$ where $\mathbf{k'}_1=\mathbf{k}_1+\mathbf{q}$ and $\mathbf{k'}_2=\mathbf{k}_2-\mathbf{q}$. Thus for $i\neq j$ the photon momentum $\mathbf{q}$ is constrained to $\mathbf{q}=\mathbf{k}%
_{1}^{\prime }-\mathbf{k}_{1}=\mathbf{k}_{2}-\mathbf{k}_{2}^{\prime }$.
However, there is \textit{ no} restriction for $n$ in the case of an ideal microcavity without loss. The fact that the momentum $\mathbf{q}$ of the virtual photon is restricted implies that it is not necessary to introduce a cutoff, but simply to consider $\mathbf{q}$ values within the range of the long-wave approximation of the electrons. 
The discretization in the $z$ direction implies that the photon propagator of the last equation can be interpreted as a sum of individual massive fields with mass $n\pi /L$. The behavior of entanglement harvesting with the field's mass has been studied in detail in Ref. \cite{gap}, where it was shown that massive fields are not interesting due to the fact that the field correlations decay with its mass, which is a result obtained in Eq. (A6) of \cite{jsa2018}. However, entanglement harvesting can increase with field mass by decreasing the noise experienced by detectors \cite{gap}. 

In free space, since there are no cavity walls imposing boundary conditions, the detectors are not excited when there is a perfect match between the energy gap and the field frequency in the monochromatic limit. This behavior is not expected in cavities, and the probability does not decrease in the resonant limit. In fact, it has been shown that the excitation probability is maximized when we take the monochromatic limit for
any number of spatial dimensions in cavities \cite{click}.
In turn, although the momenta $\mathbf{q}$ of the photon is restricted due to momentum conservation, there is no restriction to the sum of the discrete momentum perpendicular to the layers, requiring the consideration of a cutoff when real microcavities are modeled. This notable behavior is due to the fact that, although electrons are confined to live on a two-dimensional surface, the electromagnetic field among the layers lives in a confined three-dimensional space. The dynamics of the electromagnetic field in the orthogonal dimension to the layers must be traced out. 

\subsection{Temporal Taylor series approximation}
Since our only interest is studying the formation of entanglement, we consider small times in the perturbative expansion of $\rho\left(t\right)$. Then, expanding in Taylor series in $t$, the reduced density operator can be written as
\begin{equation}
	\rho\left(t\right)=\rho_{S}+\left.\frac{d\rho}{dt}\right|_{t=0}t+\frac{1}{2}\left.\frac{d^{2}\rho}{dt^{2}}\right|_{t=0}t^{2}+\mathcal{O}\left(t^{3}\right).
\end{equation}
To condense the notation we shall write the matrix elements as $\left\langle \mathbf{k}_{1}^{\prime},\mathbf{k}_{2}^{\prime}\right\vert \cdots\left\vert \mathbf{k}_{1}\mathbf{,k}{}_{2}\right\rangle =\left\langle \cdots\right\rangle $.
It is not difficult to show from Eq. (\ref{aa13})
that $d\rho/dt\left(t=0\right)=0$. For the second order contribution we get
\begin{align}
	\left\langle \left.\frac{d^{2}\rho}{dt^{2}}\right|_{t=0}\right\rangle  & =-\zeta t^{2}\sum\limits _{i,j,\lambda}\Delta_{ij}\left(\rho_{S}\sigma_{\lambda}^{i}\sigma_{-\lambda}^{j}\right.\nonumber \\
	& \left.-2\sigma_{-\lambda}^{i}\rho_{S}\sigma_{\lambda}^{j}+\sigma_{\lambda}^{i}\sigma_{-\lambda}^{j}\rho_{S}\right)\tilde{\delta}_{ij}, \label{eq:rho2}
\end{align}
where $\zeta=\left(ev_{F}\gamma/\hbar\right)^{2}=e^{2}v_{F}^{2}/\hbar L\varepsilon_0 \varepsilon_r$ and
\begin{equation}
	\Delta_{ij}={\underset{n,\mathbf{q}}{\sum}}\frac{\sin\left(\pi nd_{i}/L\right)\sin\left(\pi nd_{j}/L\right)}{\omega_{n,\mathbf{q}}}\tilde{\delta}_{ij}. \label{aa15}
\end{equation}
We shall separate the matrix elements of the reduced density operator by writing 
\begin{equation}
 \rho\left(t\right)  =\rho_{S} +\rho_{noise}+\rho_{signal},
\end{equation}
where $\rho _{noise}$ is obtained from Eq. (\ref{eq:rho2}) considering $i=j$, and $\rho _{signal}$ considering $i\neq j$. For simplicity, in what follows we omit the brackets notation $\langle ... \rangle$. The $\rho _{noise}$ contribution is local to each monolayer and thus it is not involved in the flow of information between layers, unlike the term $\rho_{signal}$ which includes the interlayer photon propagator. As we are
concerned with correlations between layers for small times, the rotating wave approximation is not applied because in this case we should remove the terms in the Hamiltonian that oscillate rapidly \cite{marti}.

\section{Results and discussions}\label{sec:RandD}

In order to obtain the critical parameters at which the reduced quantum state is entangled, we first compute the background noise of each layer, which is given by taking $i=j$ in Eq. \eqref{eq:rho2}. By making the sum in $\lambda $ and using that $\{\sigma _{+},\sigma _{-}\}=I$, we get
\begin{equation}
	\rho_{noise} =-\zeta t^{2}\sum\limits _{i}\Delta_{ii}\left(\rho_{S}-\sum\limits _{\lambda}\sigma_{-\lambda}^{i}\rho_{S}\sigma_{\lambda}^{i}\right). \label{ro2}
\end{equation}
To obtain $\ensuremath{\sum\limits _{\lambda}\sigma_{-\lambda}^{i}\rho_{S}\sigma_{\lambda}^{i}}$ we need to compute, in each layer, the action of $\sigma _{\lambda
}$ on $\left\vert \nu
,s\right\rangle $ given by Eq. (\ref{3a}). To do this, we first use that $\sigma _{\lambda }\left\vert
A,s\right\rangle =\frac{(1-\lambda )}{2}\left\vert B,s\right\rangle $, $%
\sigma _{\lambda }\left\vert B,s\right\rangle =\frac{(1+\lambda )}{2}%
\left\vert A,s\right\rangle $, and express the sublattice basis $%
\{\left\vert A,s\right\rangle ,\left\vert B,s\right\rangle \}$ in terms of the band basis $%
\{\left\vert +,s\right\rangle ,\left\vert -,s\right\rangle \}$ as
\begin{align}
	\left\vert A,s\right\rangle  & =-\frac{1}{\Delta\chi}\sum\limits _{\nu}\nu\chi_{k,-\nu,s}\sqrt{1+\chi_{k,\nu,s}^{2}}\left\vert \nu,s\right\rangle ,\\
	\left\vert B,s\right\rangle  & =\frac{e^{-i\phi_{\mathbf{k}}}}{\Delta\chi}\sum\limits _{\nu}\nu\sqrt{1+\chi_{k,\nu,s}^{2}}\left\vert\nu,s\right\rangle.
\end{align}
The action of $\sigma _{\lambda }$ on $\left\vert \nu,s\right\rangle$ then reads 
\begin{align}
	\sigma_{\lambda}\left\vert \nu,s\right\rangle  & =\frac{e^{i \lambda\phi_{\mathbf{k}}}}{2\Delta\chi}\sum\limits _{\nu^{\prime}}\nu^{\prime}\sqrt{\frac{1+\chi_{k,\nu^{\prime},s}^{2}}{1+\chi_{k,\nu,s}^{2}}}\nonumber\\ 
 &\times\left[\left(1-\lambda\right) - \left(1+\lambda\right) \chi_{k,\nu,s}\chi_{k,-\nu^{\prime},s}\right]\left\vert \nu^{\prime},s\right\rangle. \label{ro4}
\end{align}
By using the initial density operator of Eq. (\ref{ro}) and last equation we then get (cf. Appendix \ref{app:A})
\begin{align}
	\sum\limits _{\lambda}\sigma_{-\lambda}^{i}\rho_{S}\sigma_{\lambda}^{i} & =\frac{1}{\Delta\chi_{i}^{2}(1+\chi_{k_{i},\nu_{i},s_{i}}^{2})}\sum\limits _{\nu^{\prime},\nu^{\prime\prime}}\nu^{\prime}\nu^{\prime\prime}\nonumber\nonumber\nonumber \\
	& \times\sqrt{\left(1+\chi_{k_{i},\nu^{\prime},s_{i}}^{2}\right)\left(1+\chi_{k_{i},\nu^{\prime\prime},s_{i}}^{2}\right)}\nonumber \\
	& \times\left(1+\chi_{k_{i},\nu_{i},s_{i}}^{2}\chi_{k_{i},-\nu^{\prime},s_{i}}\chi_{k_{i},-\nu^{\prime\prime},s_{i}}\right)\nonumber \\
	& \times\left(\left\vert \nu^{\prime},s_{i}\right\rangle \otimes\left\vert \nu_{j},s_{j}\right\rangle \right)\left(\left\langle \nu^{\prime\prime},s_{i}\right\vert \otimes\left\langle \nu_{j},s_{j}\right\vert \right), \label{ro3}
\end{align}
where $j\ne i$. The density $\rho _{noise}$ in the basis {$\{ \left\vert \nu_{1},s_{1}\right\rangle \otimes\left\vert \nu_{2},s_{2}\right\rangle,\left\vert -\nu_{1},s_{1}\right\rangle \otimes\left\vert \nu_{2},s_{2}\right\rangle, \left\vert \nu_{1},s_{1}\right\rangle \otimes\left\vert -\nu_{2},s_{2}\right\rangle,$ $\left\vert -\nu_{1},s_{1}\right\rangle \otimes\left\vert -\nu_{2},s_{2}\right\rangle \}$} can then be written in matrix form as

\begin{equation}
	\rho_{noise}=-\zeta t^{2}\left[\begin{array}{cccc}
		\sum_{i}\Delta_{ii}A_{i} & \Delta_{22}B_{2} & \Delta_{11}B_{1} & 0\\
		\Delta_{22}B_{2} & -\Delta_{22}A_{2} & 0 & 0\\
		\Delta_{11}B_{1} & 0 & -\Delta_{11}A_{1} & 0\\
		0 & 0 & 0 & 0
	\end{array}\right],\label{eq:rhonoise}
\end{equation}
where%
\begin{align}
	A_{i} & =1-\frac{1+\chi_{k_{i},\nu_{i},s_{i}}^{2}\chi_{k_{i},-\nu_{i},s_{i}}^{2}}{\Delta\chi_{i}^{2}}=\frac{2\lambda_{so}^{2}+\epsilon_{\mathbf{k}_{i}}^{2}}{2\left(\lambda_{so}^{2}+\epsilon_{\mathbf{k}_{i}}^{2}\right)},\\
	B_{i} & =\frac{1}{\Delta\chi_{i}^{2}}\sqrt{\frac{1+\chi_{k_{i},-\nu_{i},s_{i}}^{2}}{1+\chi_{k_{i},\nu_{i},s_{i}}^{2}}}\left(1+\chi_{k_{i},\nu_{i},s_{i}}^{3}\chi_{k_{i},-\nu_{i},s_{i}}\right), \label{ro3.1}
\end{align}
cf. Appendix \ref{app:B}. The non-diagonal contributions to $\rho_{noise}$ vanish for large $\epsilon_{\mathbf{k}}$, so the probability of measuring each layer in an orthogonal state decays for large electron energies. 

Proceeding as with $\rho _{noise}$, the $\rho _{signal}$ contribution can be written as (cf. Appendix \ref{app:C})
\begin{equation}
	\rho_{signal}=\zeta t^{2}\frac{\Delta_{12}\nu_{1}\nu_{2}}{\Delta\chi_{1}\Delta\chi_{2}}\left[\begin{array}{cccc}
		0 & 0 & 0 & M^{\ast}\\
		0 & 0 & N^{\ast} & 0\\
		0 & N & 0 & 0\\
		M & 0 & 0 & 0
	\end{array}\right],
\end{equation}
where the non-diagonal elements read 
\begin{align}
	M & =\sqrt{\frac{\left(1+\chi_{k_{1},-\nu_{1},s_{1}}^{2}\right)\left(1+\chi_{k_{2},-\nu_{2},s_{2}}^{2}\right)}{\left(1+\chi_{k_{1},\nu_{1},s_{1}}^{2}\right)\left(1+\chi_{k_{2},\nu_{2},s_{2}}^{2}\right)}}\nonumber \\
	& \times\left[\chi_{k_{1},\nu_{1},s_{1}}^{2}e^{-i\left(\phi_{\mathbf{k}_{1}}-\phi_{\mathbf{k}_{2}}\right)}+\chi_{k_{2},\nu_{2},s_{2}}^{2}e^{i\left(\phi_{\mathbf{k}_{1}}-\phi_{\mathbf{k}_{2}}\right)}\right],\nonumber \\
	N & =\sqrt{\frac{\left(1+\chi_{k_{1},-\nu_{1},s_{1}}^{2}\right)\left(1+\chi_{k_{2},-\nu_{2},s_{2}}^{2}\right)}{\left(1+\chi_{k_{1},\nu_{1},s_{1}}^{2}\right)\left(1+\chi_{k_{2},\nu_{2},s_{2}}^{2}\right)}}\nonumber \\
	& \times\left[e^{-i\left(\phi_{\mathbf{k}_{1}}-\phi_{\mathbf{k}_{2}}\right)}+\chi_{k_{1},\nu_{1},s_{1}}^{2}\chi_{k_{2},\nu_{2},s_{2}}^{2}e^{i\left(\phi_{\mathbf{k}{}_{1}}-\phi_{\mathbf{k}_{2}}\right)}\right]. \label{M}
\end{align}
Finally, collecting all the terms, the density operator in momentum space reads
\begin{equation}
	\rho\left(t\right)=\left[\begin{array}{cccc}
		\tilde{\delta}_{ii}-\mathcal{L}_{1}-\mathcal{L}_{2} & \mathcal{B}_{2} & \mathcal{B}_{1} & \mathcal{M}^{\ast}\\
		\mathcal{B}_{2} & \mathcal{L}_{2} & \mathcal{N}^{\ast} & 0\\
		\mathcal{B}_{1} & \mathcal{N} & \mathcal{L}_{1} & 0\\
		\mathcal{M} & 0 & 0 & 0
	\end{array}\right], \label{rd10.1}
\end{equation} 
where 
\begin{align}
	\mathcal{L}_{i} & =\zeta t^{2}\Delta_{ii}A_{i}\,\quad\mathcal{B}_{i}=-\zeta t^{2}\Delta_{ii}B_{i}\, \label{rd11.A}\\
	\mathcal{M} & =\zeta t^{2}\frac{\Delta_{12}\nu_{1}\nu_{2}}{\Delta\chi_{1}\Delta\chi_{2}}M,\quad \mathcal{N}=\zeta t^{2}\frac{\Delta_{12}\nu_{1}\nu_{2}}{\Delta\chi_{1}\Delta\chi_{2}}N. \label{rd11.B}
\end{align} 
Although the photon propagator has no infrared divergence for $\mathbf{q}\rightarrow
0 $, the mode sum over $\Delta _{ii}$ does not converge. This should not
constrain the calculation because a real coupling between the layers and the microcavity field is frequency dependent, so the cavity should become transparent to all the frequencies above certain threshold. This ultraviolet cutoff in the orthogonal modes may alter causality. 

In Ref. \cite{brown} it is shown that the vanishing of the causality violation decreases polynomially fast as the UV cutoff is relaxed in 1+1 dimensional setups and for a discrete set of field modes. In the pointlike limit, the model is fully causal, and the signaling estimator vanishes when the detectors are spacelike separated. For smeared detectors, independently of the dimensionality of spacetime, it
has been shown that it is possible to recover an approximate causal model between two detectors for smearing $\sigma\ll L$. In this work we consider, for simplicity, a cutoff $n_{\max }$ in $\Delta _{ii}$ and $%
\Delta _{ij}$ and compute entanglement formation as a function of it. The case $n_{\max }=1$, where only the fundamental mode is taken into account, is important for those cavities with high volume compression \cite{keller,escudero0}.

The time-evolved density matrix of Eq. (\ref{rd10.1}) reduces to an $X$-state for electron momentum large compared to the Fermi momentum, at which the $B_{i}$ matrix elements vanish. The states $X$ refer to those states in which multiple matrix elements are zero ($\rho _{12}=\rho _{13}=\rho _{24}=\rho
_{34}=0$) \cite{tuyu}. Interestingly, several well-known families of states follow the $X$ form, including the Bell states, Werner states, and isotropic states \cite{wer}. Recent numerical studies have shown that all mixed two-qubit states can be transformed into $X$ states through a single unitary transformation that preserves entanglement. Consequently, the concurrence and other entanglement measures of such an $X$ state remain unchanged.

\subsection{Concurrence and entanglement entropy}

In general, a density matrix is considered non-separable or entangled when it cannot be expressed as a convex sum of local density matrices. For two-layer systems, there are different methods to parameterize the entanglement, such as concurrence, negativity and entanglement entropy {\cite{horo}}. In this work we will consider the concurrence, which for an arbitrary two-qubit state $\rho $ is equal to \cite{woo}
\begin{equation}
	\mathcal{C}\left(\rho\right)=\max\left(0,\lambda_{1}-\lambda_{2}-\lambda_{3}-\lambda_{4}\right),\label{concu0}
\end{equation}
where the $\lambda _{i}$ are the square roots of the eigenvalues of the matrix $\rho (\sigma _{y}\otimes \sigma _{y})\rho ^{\ast }(\sigma
_{y}\otimes \sigma _{y})$, ordered such that $\lambda _{1}\geq \lambda _{2}\geq \lambda _{3}\geq \lambda _{4}$. From the concurrence measure it is possible to obtain the entanglement entropy of a
two-qubit state as {\cite{woo2}}
\begin{equation}
	H\left(x\right)=-x\log_{2}\left(x\right)-\left(1-x\right)\log_{2}\left(1-x\right),\label{concu3}
\end{equation}
where $x=\frac{1}{2}\left(1+\sqrt{1-\mathcal{C}^{2}}\right)$. The entropy measures the indeterminacy of the state of the subsystems; it is a positive quantity which attains its maximum possible value $H=1$ when the state is maximally entangled at $x=0.5$.

\begin{figure*}[t]
	\includegraphics[scale=0.39]{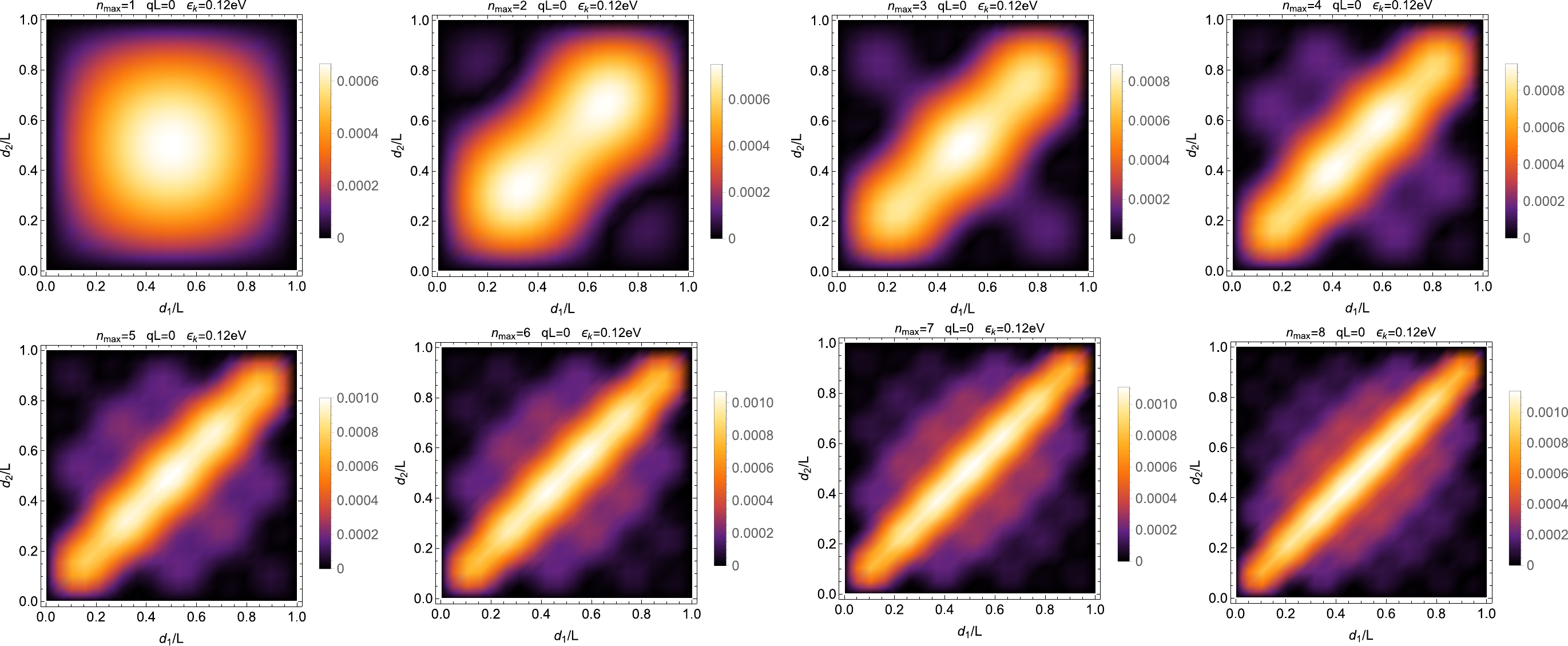}
	\caption{Concurrence over $\zeta t^{2} \lambda_{so} $ for $qL=0$ as a function of $d_{1}/L$ and $d_{2}/L$, for different
		values of cutoff $n_{\max }$ using $\epsilon_{k}=0.12$ eV, $\lambda_{so}=3.9$ meV, $\Delta\phi=0$, and $\nu_{1}=\nu_{2}= s_{1}=s_{2}=1$. }
	\label{d1d2q0}
\end{figure*}

From Eq. \eqref{rd10.1} the eigenvalues of $\rho (\sigma _{y}\otimes \sigma _{y})\rho ^{\ast }(\sigma _{y}\otimes
\sigma _{y})$ are
$\left\vert \mathcal{M}\right\vert ^{2}$ (doubly-degenerate) and $(\sqrt{\mathcal{L}_{1}\mathcal{L}_{2}} \pm \left\vert \mathcal{N}\right\vert)^{2}$.
It can be verified numerically that for realistic energies one always has $(\sqrt{\mathcal{L}_{1}\mathcal{L}_{2}} \pm \left\vert \mathcal{N}\right\vert)^{2}> \left\vert \mathcal{M}\right\vert ^{2}$. Therefore, the factor \(\sqrt{\mathcal{L}_{1}\mathcal{L}_{ 2}}\) cancels out in the computation of Eq. \eqref{concu0}, and the concurrence becomes
\begin{equation}
	\mathcal{C}=2\left(\left\vert \mathcal{N}\right\vert -\left\vert \mathcal{M}\right\vert \right).\label{concu1}
\end{equation}
Thus the concurrence only depends on the signal contribution to the density operator, i.e., is influenced only by interlayer scattering processes. From Eq. \eqref{rd10.1} we further see that the concurrence depends directly on the interlayer photon propagator. In the particular case of identical initial energies in both layers, the concurrence reduces to
\begin{align}
\mathcal{C} & =\frac{1}{2}\left\vert \nu_{1}s_{1}+\nu_{2}s_{2}\right\vert \frac{\zeta t^{2}}{\epsilon_{\mathbf{k}}^{2}+\lambda_{so}^{2}}\left\vert \Delta_{12}\right\vert \nonumber\nonumber \\
 & \times\left[\sqrt{\epsilon_{\mathbf{k}}^{4}\cos^{2}\left(\Delta\phi\right)+4\lambda_{so}^{2}\left(\epsilon_{\mathbf{k}}^{2}+\lambda_{so}^{2}\right)}-\epsilon_{\mathbf{k}}^{2}\left|\cos\left(\Delta\phi\right)\right|\right].\label{concu2}
\end{align}
This expression reflects a strong dependence with the relative angle $\Delta\phi=\phi_{\mathbf{k}_{1}}-\phi_{\mathbf{k}_{2}}$ of electrons in each layer. The prefactor $\left\vert \nu_{1}s_{1}+\nu_{2}s_{2}\right\vert$ gives non-vanishing concurrence only for the same bands or the same spins, or $\nu_{1}=-\nu_{2}=1$ and $s_{1}=-s_{2}=1$. 

\begin{figure}[t]	
	\includegraphics[scale=0.46]{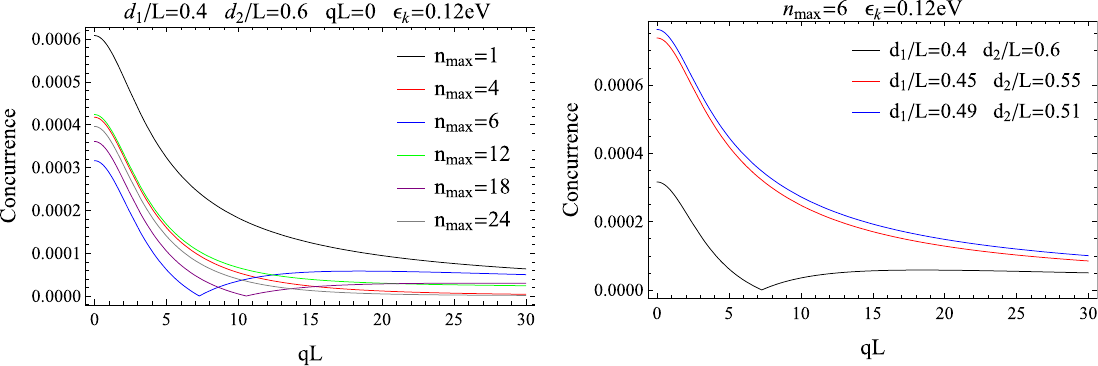}
	\caption{(Left) Concurrence as a function of momentum $q$ for different cutoffs $n_{\max }$ and particular values of $d_{1}/L$ and $d_{2}/L$. (Right) Concurrence for different values of $d_{1}/L$ and $d_{2}/L$, for cutoff $n_{\max }=1$. In both figures $\lambda_{so}=3.9$ meV, $\Delta\phi=0$, and $\nu_{1}=\nu_{2}=s_{1}=s_{2}=1$.}	\label{concuQ}
\end{figure}

\begin{figure}[t]	
	\includegraphics[scale=0.45]{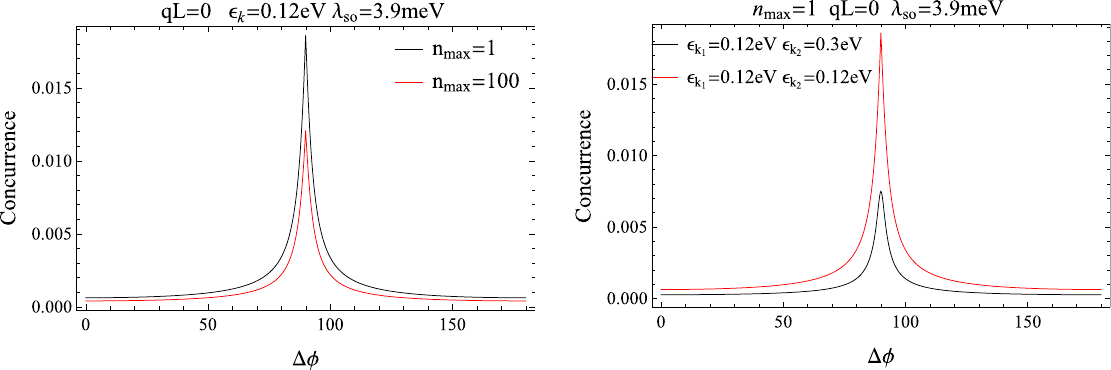}
	\caption{(Left) Concurrence as a function of the relative angle $\Delta\phi$ for different cutoffs $n_{\max }$. (Right) Concurrence as a function of the relative angle $\Delta\phi$ for different electron energies. In all the figures $d_{1}/L=0.4$, $d_{2}/L=0.6$, $\lambda_{so}=3.9\,meV$, $\nu_{1}=\nu_{2}=s_{1}=s_{2}=1$.}
	\label{concuang3}
\end{figure}

In Figure \ref{d1d2q0} we show the scaled concurrence $\mathcal{C}/ \zeta t^{2} \lambda_{so}$ as a function of $d_{i}/L$ and different mode cutoffs $n_{\max }$, for $q=0$ and identical energies $\epsilon_{\mathbf{k}}$. As $n_{\max }$ increases, the oscillatory profile of the propagators is enhanced. This behavior is more prominent for small interlayer separations $d_{2}\sim d_{1}$, where the maxima and minima of the interlayer propagator $\Delta_{12}$ follows the superposition of different terms of the form $\sim \sin ^{2}(n\pi d_{i}/L)$. By replacing $d_{2}/L=1-d_{1}/L$, and plotting the concurrence as a function of $d_{1}/L$, one further sees  that the concurrence narrows towards $\mathcal{C}=1$ around $d_{1}=L/2$ as $n_{\max }$ increases.
In Figure \ref{concuQ}, the concurrence is shown as a function of $q$ for different discrete cutoff and layer positions. Interestingly, for certain $q$ values where $\Delta_{12}$ vanishes, the time-evolved quantum state becomes separable. Note that such vanishing only occurs when many harmonics are taken into account in the propagator, so that their relative contribution can interfere destructively at certain values of $q$. In Figure \ref{concuang3}, the concurrence is shown as a function of $\Delta\phi$ for different discrete cutoffs and different electron energies. The concurrence is larger for $\Delta\phi=\pi/2$, which implies entanglement formation for perpendicular scattering between electrons. In that case Eq. \eqref{concu2} gives
\begin{equation}
\mathcal{C}_{\mathrm{max}}=\zeta t^{2}\frac{\lambda_{so}}{\sqrt{\epsilon_{\mathbf{k}}^{2}+\lambda_{so}^{2}}}\left\vert \Delta_{12}\right\vert .
\end{equation}
Thus in the regime $\epsilon_{\mathbf{k}}\gg \lambda_{so}$ the concurrence maximum scales as $\mathcal{C}_{\mathrm{max}}\sim \lambda_{so}/\epsilon_{\mathbf{k}}$, so high SOI and low energies favor the entanglement formation.

\begin{figure}[t]
	\includegraphics[scale=0.46]{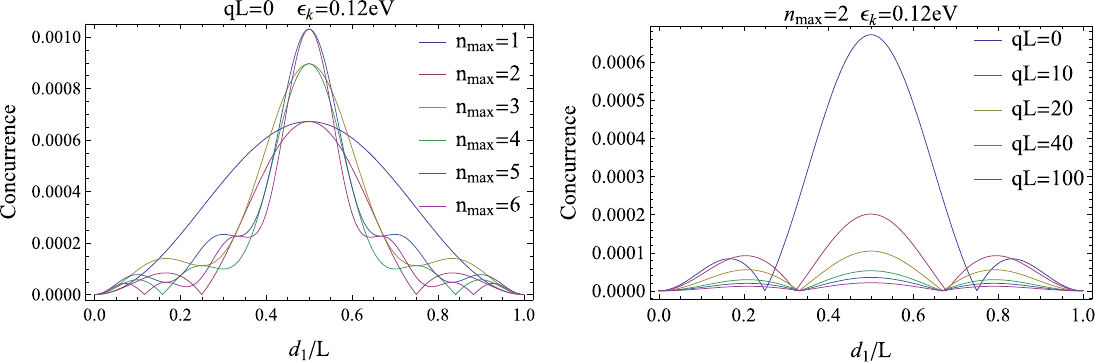}
	\caption{Concurrence as a function of $d_{1}/L$ for different $n_{\max }$ (left) and $qL$ (right), showing the minima and maxima intensity of entanglement formation In both figures $\lambda_{so}=3.9$ meV, $\Delta\phi=0$,  $\nu_{1}=\nu_{2}= s_{1}=s_{2}=1$.}
	\label{concu1d}
\end{figure}

In Figure \ref{concu1d} we see that the maximum entanglement always occurs when the layers are close to the center of the cavity, that is, $d_1/L\sim d_2/L\sim1/2$. This is expected because of the dominant contribution of the fundamental cavity mode. The values at which the concurrence vanishes shift as $n_{max}$ and $qL$ vary due to the non-trivial dependence of the photon propagator $\Delta_{12}$ with these parameters, cf. Eq. \eqref{aa15}. 

\begin{figure}[t]
	\includegraphics[scale=0.46]{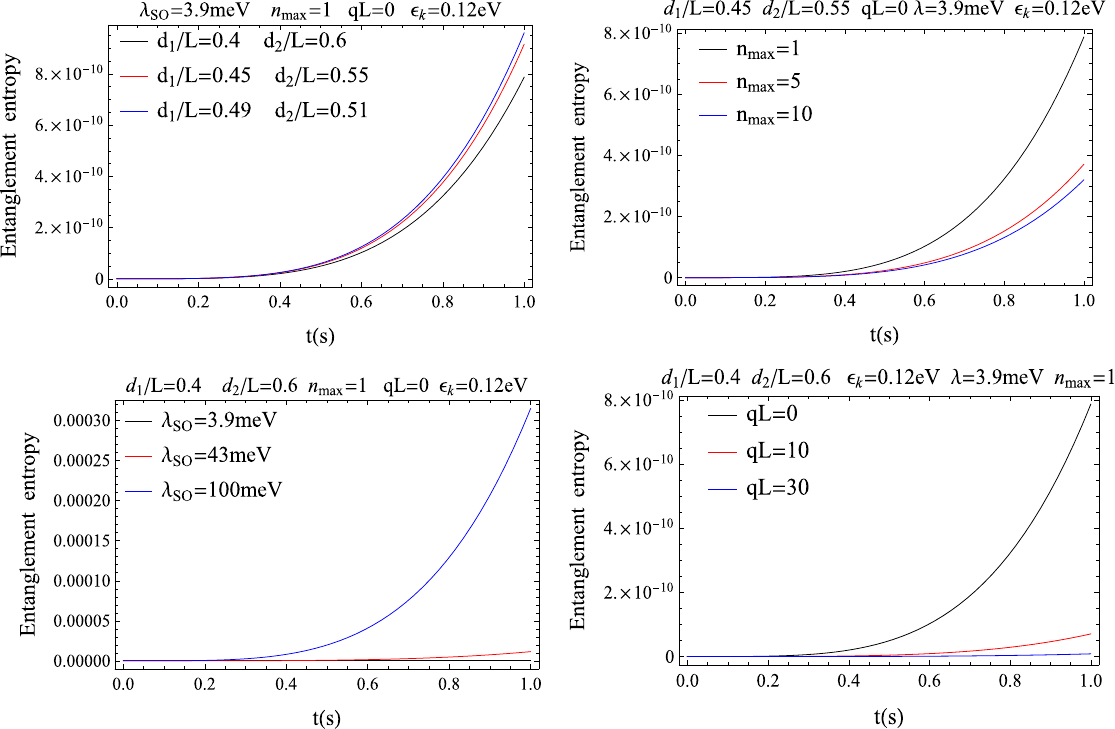}
	\caption{Entanglement entropy for different set of parameters as a function of time, for the case $\nu_{1}=\nu_{2}= s_{1}=s_{2}=1$ and different layer positions, cavity modes, SOI interaction and in-plane exchanged momentum.}
	\label{entang}
\end{figure}

The entanglement entropy is represented over time in Figure \ref{entang}, where we have considered the vacuum permittivity. The introduction of a dielectric in the cavity alters the refractive index and decreases the coupling parameter $\gamma$ \cite{bada}. In agreement with the concurrence behavior, the entanglement entropy is observed to be higher for smaller interlayer distances and decreases with a large discrete cutoff $n_{\max}$. Notably, in materials like stanene where the spin-orbit coupling is relatively large ($\lambda_{so}\sim0.1\,\mathrm{eV}$), the entanglement entropy increases. Additionally, the entanglement is favored for small photon momentum transfer, suggesting suitability for electronic quantum correlations mediated by the quantized photon modes in the microcavity.

The entanglement formation increases, in leading order, as a function of time as $\propto t^{2}$. As this behavior follows from a Taylor expansion of the density matrix around $t=0$, our results are technically valid at small times; in particular, for times smaller that the light-crossing time between the layers. Thus, the correlation between the layers can be non-casual, so the entanglement formation is expected to come mainly from a genuinely harvested correlation from the field \cite{tjoa2}. We note that in Ref. \cite{facu} a similar protocol to the one given in Ref. \cite{tjoa2} was introduced, where correlations are non-causal for space-like separations of the layers. 

Our results show that buckled honeycomb lattices placed inside a microcavity are susceptible to be entangled through the exchange of virtual photons and that this effect is strongly dependent on the spin-orbit coupling and the dynamical parameters of the electrons. Previous works have shown that entanglement harvesting also depends on the internal structure of the detectors, which in the most simple case is defined by the detector gaps. It is source of future work to study the possible formation of non-causal correlations between different monolayers inside the microcavity, which allows to emulate different gaps through the different SOI parameters. In particular, the application of an external electric field in a buckled two-dimensional lattice leads to a spin-dependent gap that allows one to study momentum-spin entanglement as done in Refs. \cite{lamata} and \cite{pacho}.

\section{Conclusions}\label{sec:Conclusions}

We have studied the entanglement harvesting between two buckled honeycomb lattices placed inside a microcavity, induced by the vacuum state of the electromagnetic field. Applying time-dependent perturbation theory and partially tracing out the cavity field degrees of freedom, we computed the concurrence as a measure of entanglement formation and analyzed the set of parameters involved in the entanglement harvesting. We have shown that the concurrence depends linearly on the cavity-mediated interlayer interaction, so that the entanglement region follows the spatial profile of the photon propagator. The magnitude of the concurrence depends directly on the spin-orbit coupling strength, reflecting that a buckled geometry is favorable for entanglement formation. The entanglement zones also depend strongly on the momentum of the exchanged photons, with minimum values appearing when the interlayer photon propagator vanishes. 

\section{Acknowledgment}

This article was partially supported by CONICET grants (Argentina National Research Council) and Universidad Nacional del Sur (UNS) and ANPCyT through PICT 2019- 03491 Res. No. 015/2021, and PIP-CONICET 2021-2023 Grant
No. 11220200100941CO. J.S.A. and A.J. acknowledge support as members of CONICET, F.E. acknowledges support from a research fellowship from this institution, and F.A. acknowledges support as a member of Departamento de F%
\'{\i}sica, Universidad Nacional del Sur.

\appendix

\bigskip
\onecolumngrid
\section{Trace of photon degrees of freedom}\label{app:A}
From Eq. \eqref{a9}, up to second order we can write the density matrix as $  \rho_T = \rho_0 + \rho_1 + \rho_2$. The partial trace over the cavity field degrees of freedom is computed as
\begin{equation}
T_{r}\left(\rho_{T}\right)=\left\langle \Omega_{0}\right\vert \rho_{T}\left\vert \Omega_{0}\right\rangle +\sum\limits_{n,\mathbf{q},\lambda}\left\langle 1_{n,\mathbf{q},\lambda}\right\vert \rho_{T}\left\vert 1_{n,\mathbf{q},\lambda}\right\rangle +\sum\limits_{n,\mathbf{q},\lambda;n^{\prime},\mathbf{q}^{\prime},\lambda^{\prime}}\left\langle 1_{n,\mathbf{q},\lambda},1_{n^{\prime},\mathbf{q}^{\prime},\lambda^{\prime}}\right\vert \rho_{T}\left\vert 1_{n,\mathbf{q},\lambda},1_{n^{\prime},\mathbf{q}^{\prime},\lambda^{\prime}}\right\rangle +...,\label{A1}
\end{equation}
where $\left\vert \Omega_0\right\rangle $ is the ground state, $\left\vert 1_{n,\mathbf{q},\lambda} \right\rangle$ is the quantum state of the photon field with one excitation, $\left\vert 1_{n,\mathbf{q},\lambda}, 1_{n^{\prime},\mathbf{q}^{\prime},\lambda^{\prime}}\right\rangle$ with two excitations, and so on. From Eq. \eqref{a11} it can be  easily checked that at zero order $T_r (\rho_0) = \rho_S$. Furthermore, since the first-order evolution operator $U_1$ depends linearly with the potential vector $A_{\lambda }^{i}$ that creates
or annihilates a cavity quantum, from Eq. \eqref{A2} it follows that $T_r(\rho_1)=0$.

At second order, the contribution $\rho
_{2}=U_{2}\rho _{0}+U_{1}\rho _{0}U_{1}^{\dagger }+\rho_{0}U_{2}^{\dagger}$ involves the product of two evolution operators $U_1$, and one second-order operator $U_2$. This generally yields two products on $\rho_0$, which contain terms such as $a_{n,\mathbf{q},\lambda}a_{n^{\prime},\mathbf{q}^{\prime},\lambda^{\prime}}^{\dagger}$ that gives 
$U_{2}\rho_{0}\sim a_{n,\mathbf{q},\lambda}\left\vert 1_{n^{\prime},\mathbf{q}^{\prime},-\lambda^{\prime}}\right\rangle \left\langle \Omega\right\vert =\delta_{n,n^{\prime}}\delta_{\mathbf{q},\mathbf{q}^{\prime}}\delta_{\lambda,-\lambda^{\prime}}\left\vert \Omega\right\rangle \left\langle \Omega\right\vert $. Using this and restoring all the
parameters, the partial trace of the contribution $\sim U_2\rho_0$ reads
\begin{equation}
T_r(U_{2}\rho _{0})=-ev_{F}\sum\limits_{i,j;\lambda ,\lambda ^{\prime
}}^{{}}\sum\limits_{n,\mathbf{q}}^{{}}\int_{t_{0}}^{t}%
\int_{t_{0}}^{t}dt_{1}dt_{2}\left\langle \Omega \right\vert A_{\lambda
}^{i}(\mathbf{r}_{i},t_{i})A_{\lambda ^{\prime }}^{j}(\mathbf{r}%
_{j},t_{j})\left\vert \Omega \right\rangle \sigma _{\lambda
}^{i}(t_{1})\sigma _{\lambda ^{\prime }}^{j}(t_{2})\rho _{S},
\label{A2}
\end{equation}
where $\left\langle \Omega \right\vert A_{\lambda }^{i}(\mathbf{r}%
_{i},t_{i})A_{\lambda ^{\prime }}^{j}(\mathbf{r}_{j},t_{j})\left\vert \Omega
\right\rangle $ is the photon propagator. The same analysis can be performed with the contribution $T_{r}\left(\rho_{0}U_{2}^{\dagger}\right)$. The other contribution $U_{1}\rho _{0}U_{1}^{\dagger }$ to $\rho_2$ involves the product $%
A_{\lambda }^{i}\left\vert \Omega \right\rangle \left\langle \Omega
\right\vert A_{\lambda ^{\prime }}^{j}$ which implies that when the partial
trace is computed, those terms in which the final state $A_{\lambda
}^{i}\left\vert \Omega \right\rangle $ is identical to the final state $%
\left\langle \Omega \right\vert A_{\lambda ^{\prime }}^{j}$ will survive.
It is then not difficult to show that the unique term that contributes to the
partial trace of $U_{1}\rho _{0}U_{1}^{\dagger }$ reduces to the last term in Eq. \eqref{aa13}. A different perspective of the same calculation is noting
that $T_{r}\left(U_{1}\rho_{0}U_{1}^{\dagger}\right)=T_{r}\left(U_{1}\left\vert \Omega\right\rangle \left\langle \Omega\right\vert \rho_{S}U_{1}^{\dagger}\right)=T_{r}\left(\left\langle \Omega\right\vert U_{1}^{\dagger}U_{1}\left\vert \Omega\right\rangle \right)\rho_{S}=\left\langle \Omega\right\vert U_{1}^{\dagger}U_{1}\left\vert \Omega\right\rangle \rho_{S}$.  

\section{Matrix elements of $ \rho_{noise}$}\label{app:B}

The matrix elements of $\rho_{noise}$, given by Eq. \eqref{eq:rhonoise}, follow by summing over the indices $\nu',\nu''=\nu_{i}$ in Eq. \eqref{ro3}, and then projecting Eq. \eqref{ro2} over the basis $\ensuremath{\{\left\vert \nu_{1},s_{1}\right\rangle \otimes\left\vert \nu_{2},s_{2}\right\rangle ,\left\vert -\nu_{1},s_{1}\right\rangle \otimes\left\vert \nu_{2},s_{2}\right\rangle ,\left\vert -\nu_{1},s_{1}\right\rangle \otimes\left\vert \nu_{2},s_{2}\right\rangle ,}\ensuremath{\left\vert -\nu_{1},s_{1}\right\rangle \otimes\left\vert -\nu_{2},s_{2}\right\rangle \}}$.
Explicitly, from Eq. \eqref{ro3} we have
\begin{align}
\sum\limits_{\lambda}\sigma_{-\lambda}^{i}\rho_{S}\sigma_{\lambda}^{i} & =\frac{1+\chi_{k_{i},\nu_{i},s_{i}}^{2}\chi_{k_{i},-\nu_{i},s_{i}}^{2}}{\Delta\chi_{i}^{2}}\left\vert \nu_{i},s_{i}\right\rangle \otimes\left\vert \nu_{j},s_{j}\right\rangle \left\langle \nu_{i},s_{i}\right\vert \otimes\left\langle \nu_{j},s_{j}\right\vert \nonumber \\
 & -\frac{1}{\Delta\chi_{i}^{2}}\sqrt{\frac{1+\chi_{k_{i},-\nu_{i},s_{i}}^{2}}{1+\chi_{k_{i},\nu_{i},s_{i}}^{2}}}\left(1+\chi_{k_{i},\nu_{i},s_{i}}^{3}\chi_{k_{i},-\nu_{i},s_{i}}\right)\left\vert \nu_{i},s_{i}\right\rangle \otimes\left\vert \nu_{j},s_{j}\right\rangle \left\langle -\nu_{i},s_{i}\right\vert \otimes\left\langle \nu_{j},s_{j}\right\vert \nonumber \\
 & -\frac{1}{\Delta\chi_{i}^{2}}\sqrt{\frac{1+\chi_{k_{i},-\nu_{i},s_{i}}^{2}}{1+\chi_{k_{i},\nu_{i},s_{i}}^{2}}}\left(1+\chi_{k_{i},\nu_{i},s_{i}}^{3}\chi_{k_{i},-\nu_{i},s_{i}}\right)\left\vert -\nu_{i},s_{i}\right\rangle \otimes\left\vert \nu_{j},s_{j}\right\rangle \left\langle \nu_{i},s_{i}\right\vert \otimes\left\langle \nu_{j},s_{j}\right\vert \nonumber \\
 & +\frac{\sqrt{1+\chi_{k_{i},-\nu_{i},s_{i}}^{2}}\left(1+\chi_{k_{i},\nu_{i},s_{i}}^{4}\right)}{\Delta\chi_{i}^{2}(1+\chi_{k_{i},\nu_{i},s_{i}}^{2})}\left\vert -\nu_{i},s_{i}\right\rangle \otimes\left\vert \nu_{j},s_{j}\right\rangle \left\langle -\nu_{i},s_{i}\right\vert \otimes\left\langle \nu_{j},s_{j}\right\vert ,
\end{align}
where we have used that $\nu_{i}^{2}=1$. The contribution of last equation to the first matrix element $\rho_{noise}$ then reads
\begin{equation}
\left\langle \nu_{i},s_{i}\right\vert \otimes\left\langle \nu_{j},s_{j}\right\vert \left(\sum\limits_{\lambda}\sigma_{-\lambda}^{i}\rho_{S}\sigma_{\lambda}^{i} \right)\left\vert \nu_{i},s_{i}\right\rangle \otimes\left\vert \nu_{j},s_{j}\right\rangle =\frac{1+\chi_{k_{i},\nu_{i},s_{i}}^{2}\chi_{k_{i},-\nu_{i},s_{i}}^{2}}{\Delta\chi_{i}^{2}}.
\end{equation}
Now, from Eq. \eqref{eq:chi} it follows that
\begin{align}
\chi_{k,\nu,s}^{2} & =\frac{1}{\epsilon_{\mathbf{k}}^{2}}\left(\nu_{i}\sqrt{\lambda^{2}+\epsilon_{\mathbf{k}}^{2}}-s\lambda\right)^{2},\\
\Delta\chi^{2}= & \left(\chi_{k,+,s}-\chi_{k,-,s}\right)^{2}=\frac{4\left(\lambda^{2}+\epsilon_{\mathbf{k}}^{2}\right)}{\epsilon_{\mathbf{k}}^{2}},
\end{align} \label{imp}
and thus
\begin{equation}
\frac{1+\chi_{k_{i},\nu_{i},s_{i}}^{2}\chi_{k_{i},-\nu_{i},s_{i}}^{2}}{\Delta\chi_{i}^{2}}=\frac{\epsilon_{\mathbf{k}_{i}}^{2}}{2\left(\lambda^{2}+\epsilon_{\mathbf{k}_{i}}^{2}\right)}.
\end{equation}
Since $\left\langle \nu_{i},s_{i}\right\vert \otimes\left\langle \nu_{j},s_{j}\right\vert \rho_{S}\left\vert \nu_{i},s_{i}\right\rangle \otimes\left\vert \nu_{j},s_{j}\right\rangle =1$,
from Eq. \eqref{ro2} we then have
\begin{equation}
\left\langle \nu_{i},s_{i}\right\vert \otimes\left\langle \nu_{j},s_{j}\right\vert \rho_{noise}\left\vert \nu_{i},s_{i}\right\rangle \otimes\left\vert \nu_{j},s_{j}\right\rangle =-\zeta t^{2}\sum\limits_{i}\Delta_{ii}\left[1-\frac{\epsilon_{\mathbf{k}_{i}}^{2}}{2\left(\lambda^{2}+\epsilon_{\mathbf{k}_{i}}^{2}\right)}\right],
\end{equation}
which reduces to the first matrix element in Eq. \eqref{eq:rhonoise}. The same procedure yields all the other matrix elements of $\rho_{noise}$.

\section{Matrix elements of $\rho_{signal}$}\label{app:C}

The contribution $\rho _{signal}$ is obtained from Eq. \eqref {eq:rho2} taking $i\neq j$:
\begin{equation}
\rho_{signal}=-\zeta t^{2}\sum\limits_{i\neq j,\lambda}\Delta_{ij}\left(\sigma_{\lambda}^{i}\sigma_{-\lambda}^{j}\rho_{S}+\rho_{S}\sigma_{\lambda}^{i}\sigma_{-\lambda}^{j}-2\sigma_{-\lambda}^{j}\rho_{S}\sigma_{\lambda}^{i}\right).
\end{equation}
Using Eq. \eqref{ro4} we then get
\begin{align}
\rho_{signal} & =-\frac{\zeta t^{2}\Delta_{12}}{\Delta\chi_{1}\Delta\chi_{2}\sqrt{1+\chi_{k_{1},\nu_{1},s_{1}}^{2}}\sqrt{1+\chi_{k_{2},\nu_{2},s_{2}}^{2}}}\sum\limits_{\nu_{1}^{\prime}\nu_{2}^{\prime}}\nu_{1}^{\prime}\nu_{2}^{\prime}\sqrt{1+\chi_{k_{1},\nu_{1}^{\prime},s_{1}}^{2}}\sqrt{1+\chi_{k_{2},\nu_{2}^{\prime},s_{2}}^{2}}\nonumber \\
 & \times\left[\left(\chi_{k_{1},\nu_{1},s_{1}}\chi_{k_{1},-\nu_{1}^{\prime},s_{1}}e^{-i\phi_{\mathbf{k}_{1}}}e^{i\phi_{\mathbf{k}_{2}}}+\chi_{k_{2},\nu_{2},s_{2}}\chi_{k_{2},-\nu_{2}^{\prime},s_{2}}e^{i\phi_{\mathbf{k}_{1}}}e^{-i\phi_{\mathbf{k}_{2}}}\right)\left\vert \nu_{1}^{\prime},s_{1}\right\rangle \left\langle \nu_{1},s_{1}\right\vert \otimes\left\vert \nu_{2}^{\prime},s_{2}\right\rangle \left\langle \nu_{2},s_{2}\right\vert \right.\nonumber \\
 & +\left(\chi_{k_{1},\nu_{1},s_{1}}\chi_{k_{1},-\nu_{1}^{\prime},s_{1}}e^{i\phi_{\mathbf{k}_{1}}}e^{-i\phi_{\mathbf{k}_{2}}}+\chi_{k_{2},\nu_{2},s_{2}}\chi_{k_{2},-\nu_{2}^{\prime},s_{2}}e^{-i\phi_{\mathbf{k}_{1}}}e^{i\phi_{\mathbf{k}_{2}}}\right)\left\vert \nu_{1},s_{1}\right\rangle \left\langle \nu_{1}^{\prime},s_{1}\right\vert \otimes\left\vert \nu_{2},s_{2}\right\rangle \left\langle \nu_{2}^{\prime},s_{2}\right\vert \nonumber \\
 & +\left(\chi_{k_{1},\nu_{1},s_{1}}\chi_{k_{2},\nu_{2},s_{2}}\chi_{k_{1},-\nu_{1}^{\prime},s_{1}}\chi_{k_{2},-\nu_{2}^{\prime},s_{2}}e^{i\phi_{\mathbf{k}_{1}}}e^{-i\phi_{\mathbf{k}_{2}}}+e^{-i\phi_{\mathbf{k}_{1}}}e^{i\phi_{\mathbf{k}_{2}}}\right)\left\vert \nu_{1},s_{1}\right\rangle \left\langle \nu_{1}^{\prime},s_{1}\right\vert \otimes\left\vert \nu_{2}^{\prime},s_{2}\right\rangle \left\langle \nu_{2},s_{2}\right\vert \nonumber \\
 & \left.+\left(\chi_{k_{1},\nu_{1},s_{1}}\chi_{k_{2},\nu_{2},s_{2}}\chi_{k_{1},-\nu_{1}^{\prime},s_{1}}\chi_{k_{2},-\nu_{2}^{\prime},s_{2}}e^{-i\phi_{\mathbf{k}_{1}}}e^{i\phi_{\mathbf{k}_{2}}}+e^{i\phi_{\mathbf{k}{}_{1}}}e^{-i\phi_{\mathbf{k}_{2}}}\right)\left\vert \nu_{1}^{\prime},s_{1}\right\rangle \left\langle \nu_{1},s_{1}\right\vert \otimes\left\vert \nu_{2},s_{2}\right\rangle \left\langle \nu_{2}^{\prime},s_{2}\right\vert \right].\label{q2}
\end{align}
For simplicity let us consider the first matrix element of $\rho _{signal}$, given by taking $\nu _{1}^{\prime }=\nu _{1}$ and $\nu
_{2}^{\prime }=\nu _{2}$,
\begin{align}
\left\langle \nu_{1},s_{1}\right\vert \otimes\left\langle \nu_{2},s_{2}\right\vert \rho_{signal}\left\vert \nu_{1},s_{1}\right\rangle \otimes\left\vert \nu_{2},s_{2}\right\rangle = & -\frac{\zeta t^{2}\Delta_{12}}{\Delta\chi_{1}\Delta\chi_{2}}2\cos\left(\phi_{\mathbf{k}_{1}}-\phi_{\mathbf{k}_{2}}\right)\left(\chi_{k_{1},\nu_{1},s_{1}}\chi_{k_{1},-\nu_{1},s_{1}}+\chi_{k_{2},\nu_{2},s_{2}}\chi_{k_{2},-\nu_{2},s_{2}}\right.\nonumber \\
 & \qquad\qquad\left.+2\chi_{k_{1},\nu_{1},s_{1}}\chi_{k_{1},-\nu_{1},s_{1}}\chi_{k_{2},\nu_{2},s_{2}}\chi_{k_{2},-\nu_{2},s_{2}}\right).
\end{align}
Using Eq. \eqref{eq:chi}, after a large but straightforward computation it can be shown that
\begin{equation}
\chi _{k_{1},\nu _{1},s_{1}}\chi _{k_{1},-\nu _{1},s_{1}}+\chi
_{k_{2},\nu _{2},s_{2}}\chi _{k_{2},-\nu _{2},s_{2}}+2\chi _{k_{1},\nu
_{1},s_{1}}\chi _{k_{1},-\nu _{1},s_{1}}\chi _{k_{2},\nu _{2},s_{2}}\chi
_{k_{2},-\nu _{2},s_{2}}=0, \label{q4}
\end{equation}%
which implies that  the first matrix element of $\rho_{signal}$ vanishes. In a similar way, the matrix element $\left\langle -\nu
_{1},s_{1}\right\vert \otimes \left\langle -\nu _{2},s_{2}\right\vert \rho
_{signal}\left\vert \nu_{1},s_{1}\right\rangle \otimes \left\vert
\nu_{2},s_{2}\right\rangle $ can be obtained from the first term of the r.h.s.
of Eq. (\ref{q2}) with $\nu _{1}^{\prime }=-\nu _{1}$ and $\nu _{2}^{\prime
}=-\nu _{2}$, which yields
\begin{align}
\left\langle -\nu_{1},s_{1}\right\vert \otimes\left\langle -\nu_{2},s_{2}\right\vert \rho_{signal}\left\vert \nu_{1},s_{1}\right\rangle \otimes\left\vert 
\nu_{2},s_{2}\right\rangle  & =\frac{\zeta t^{2}\Delta_{12}\sqrt{1+\chi_{k_{1},-\nu_{1},s_{1}}^{2}}\sqrt{1+\chi_{k_{2},-\nu_{2},s_{2}}^{2}}}{\Delta\chi_{1}\Delta\chi_{2}\sqrt{1+\chi_{k_{1},\nu_{1},s_{1}}^{2}}\sqrt{1+\chi_{k_{2},\nu_{2},s_{2}}^{2}}}\nonumber \\
 & \times\left(\chi_{k_{1},\nu_{1},s_{1}}^{2}e^{-i\phi_{\mathbf{k}_{1}}}e^{i\phi_{\mathbf{k}_{2}}}+\chi_{k_{2},\nu_{2},s_{2}}^{2}e^{i\phi_{\mathbf{k}_{1}}}e^{-i\phi_{\mathbf{k}_{2}}}\right).
\end{align}
This leads to the factor $M$ is defined in Eq. \eqref{M}. A similar procedure can be carried with the other matrix elements.

\end{document}